\documentstyle[preprint,aps,prd,eqsecnum,tighten,epsfig]{revtex} 
\begin{document}
\draft
\preprint{UMDPP\#99-015}
\title{Defect Formation and Critical Dynamics in the Early Universe}
\author{G. J. Stephens\thanks{Electronic address: 
{\tt gstephen@physics.umd.edu}} $^1$, E. A. Calzetta \thanks{Electronic address: {\tt calzetta@df.uba.ar}} $^2$,  B. L. Hu
\thanks{Electronic address: {\tt hub@physics.umd.edu}} $^1$, S. A. Ramsey \thanks{Electronic 
address: {\tt sramsey@physics.umd.edu}} $^1$}

\address{$^1$ Department of Physics, University of Maryland,
College Park, Maryland 20742-4111}

\address{$^2$ Department of Physics and IAFE, University of Buenos Aires, Argentina}
\date{\today}
\maketitle
\begin{abstract}
We study the nonequilibrium dynamics leading to the formation of topological defects 
in a symmetry-breaking phase transition of a quantum scalar field with
$\lambda \Phi^4$ self-interaction in a spatially flat, radiation-dominated Friedmann-Robertson Walker Universe.  
The quantum field is initially in a finite-temperature symmetry-restored state and the phase transition develops as the
Universe expands and cools. We present a 
first-principles, microscopic approach in which the nonperturbative, nonequilibrium dynamics of the quantum field is derived from the
two-loop, two-particle-irreducible closed-time-path effective action.  We numerically solve the dynamical 
equations for the two-point function
and we identify signatures of topological defects in the infrared portion of the momentum-space power spectrum. We find that the density of
topological defects formed after the phase transition scales as a power law with the expansion rate of the universe. We calculate the
equilibrium critical exponents of the correlation length and relaxation time for this model and show that the
power law exponent of the defect density, for both overdamped and underdamped evolution, is in good agreement with 
the ``freeze-out'' scenario proposed by Zurek.  We also introduce an analytic dynamical model, valid near the critical point, that exhibits 
the same power-law scaling of the defect density with the quench rate.  By incorporating the realistic quench of the 
expanding Universe our approach illuminates the {\it dynamical} mechanisms important for topological defect formation 
and provides a preliminary step towards a complete and 
rigorous picture of defect formation in a second-order phase transition of a quantum field. The observed power law scaling
of the defect density with the quench rate, calculated here in a quantum field theory context, provides evidence for the
``freeze-out'' scenario in three spatial dimensions. 
   
\end{abstract}
\newpage

\section{Introduction}

\subsection{Motivation}

Phase transitions in the early Universe played a decisive role in shaping 
the Universe we observe today.  In the hot early Universe it is likely that  
the broken symmetries of the standard model of particle physics were partially or totally restored in
a grand unified theory (GUT) and that a number of phase transitions occurred as the Universe expanded 
and cooled.  These include 
transitions at the GUT scale $T_c \approx 10^{14}-10^{16}$ GeV, 
at the electroweak scale $T_c \approx 10^2$ GeV and the 
color deconfinement and/or chiral phase transitions of QCD with
$T_c \approx 100$ MeV.  An inflationary phase transition, a possible explanation for the flatness, horizon
and monopole problems of the standard Friedmann-Robertson-Walker (FRW) cosmology, may 
have occurred at the GUT scale or before.  At still earlier epochs near 
the Planck energy, $T_c \approx 10^{19}$ GeV, candidate theories of
quantum gravity may allow for a phase transition to occur and give rise to the classical 
properties of spacetime described by general relativity. These first moments of the Universe contain 
clues to the solution of such outstanding problems as the origin of 
large-scale structure and matter-antimatter asymmetry and the nature 
of dark matter.  The study of the consequences of 
cosmological phase transitions may therefore hold the key to our understanding of the 
Universe. 

In a field theory with 
degenerate vacua, a symmetry-breaking phase transition can produce topological defects. 
Relics of the high-temperature symmetric phase of the theory, topological defects are 
topologically stable field configurations that are locally 
trapped in an excited state above the vacuum.  
Examples of defects include vortices in type II superconductors, superfluid 
Helium-3 and Helium-4, and cosmic strings formed in the early Universe. 
Topological defects are classified by the homotopy groups $\Pi_n(M)$ of the vacuum
manifold, $M$. In three spatial dimensions simple defects are domain 
walls if $\Pi_0(M) \neq 1$, strings if $\Pi_1(M) \neq 1$, and 
monopoles if $\Pi_2(M) \neq 1$ \cite{VilShel}. 

Kibble was the first to show that topological defects are a generic
feature of cosmological phase transitions \cite {Kib}.  In the course of  
a symmetry-breaking phase transition, as the field decays 
to the stable vacuum, it can choose the same
vacuum state on length scales only as large as the correlation length.  In the
laboratory, if the phase transition proceeds slowly, the correlation length is 
bounded only by the size of the system and the field will effectively choose
a homogeneous vacuum state.  In the early Universe, however, the correlation
length is bounded by the size of the particle horizon.  With
correlated regions limited in size by causality, Kibble argued that phase transitions in the
early
Universe necessarily leave a domain structure of vacuum states and
form topological defects.

Topological defects formed in the early Universe can have profound 
consequences
on the subsequent spacetime evolution.  In the standard FRW 
cosmology, GUT scale monopoles, even if formed with an initial 
density equal to 
Kibble's lower bound of one defect per horizon volume, contribute 
more than $10^{12}$ times the largest possible density of the Universe  
consistent with observations. The overproduction of monopoles  
is a puzzle within the standard FRW cosmology 
and provided part of the original motivation for an early epoch of 
inflation \cite{Guth}. Other topological defects such as cosmic strings are viable
candidates for the seeds of structure formation and would produce a characteristic 
signature in the angular power spectrum of the cosmic microwave background.  Satellites such as 
MAP and Plank designed specifically to measure fluctuations in the cosmic
microwave background radiation to high accuracy are scheduled to be 
launched in the near future.  The data obtained from these
missions is expected to clarify the precise role of topological defects 
in the origin of structure formation in the Universe.

The importance of topological defects as candidates for the seeds of structure
formation provides strong motivation to extract predictions from 
defect models that can be compared 
with observations.  These predictions generally contain
three main ingredients: the density of topological defects
immediately following the completion of the phase 
transition, the subsequent evolution of the defect distribution
and the calculation of physical observables such as cosmic
microwave background radiation 
anisotropies and polarization.  While both numerical and analytical 
work has been done on the evolution of the defect distribution and the
extraction of physical 
predictions, less attention has been focused on quantitative predictions of the initial defect
density which provides the starting conditions for the evolution of 
the defect distribution.     

The dynamic origins of defect formation are also themselves of 
considerable physical interest.  Formed in out-of-equilibrium processes, 
defects offer insight
into the nature of critical dynamics.  In addition, the 
ubiquitous nature of phase transitions links the study of the early
Universe with the laboratory
and allows the observation of cosmologically important mechanisms in 
condensed matter systems \cite{Zur} \cite{Volovik}. Recently, researchers
have exploited these similarities by observing defect formation in 
liquid crystals \cite{LiqCry} and superfluid Helium \cite {DefExp}. 
In either condensed matter systems or the early Universe, the density of topological defects 
immediately following the phase transition is of substantial 
physical importance. 

\subsection{The Kibble and Zurek Mechanisms of Defect Formation}

The first estimate of the initial defect density in a cosmological context was made
by Kibble \cite {Kib}.  The basic ingredients of the 
Kibble mechanism are causality and the Ginzburg temperature, $T_G$.  The 
Ginzburg temperature is defined as the temperature at which thermal fluctuations contain just enough energy for 
correlated regions of the field to overcome the potential energy barrier
between inequivalent vacua,
 
\begin {equation}
k_bT_G \sim \xi(T_G)^3 \Delta F(T_G),
\end {equation}

\noindent  where $\Delta F$ is the difference in free energy density between the true and false vacua and
 $\xi$ is the equilibrium correlation length.
In the Kibble mechanism, the length scale characterizing the initial defect network is set by the equilibrium 
correlation length of the field, evaluated at $T_G$.
In a recent series of experiments \cite {DefExp}, the Kibble mechanism was tested in the laboratory.  The results, while
confirming the production of defects in a symmetry-breaking phase transition, 
indicate that 
$\xi(T_G)$ does {\it not} set the characteristic length scale of the initial defect distribution.  These
experiments were suggested by Zurek, who criticized Kibble's
use of equilibrium arguments and the Ginzburg temperature, $T_G$ \cite {Zur}.
  
Combining equilibrium and non equilibrium ingredients, Zurek offers a
``freeze-out'' proposal to estimate the initial density of defects. 
In the Zurek proposal, above the critical temperature, the field starts off in thermal equilibrium with a heat bath.  As the 
temperature of the bath is lowered adiabatically, the field remains in local thermal equilibrium with the heat bath. Near the phase transition,  
the equilibrium correlation 
length and the equilibrium relaxation time of the field grow without bound as  

\begin {eqnarray}
\xi = \xi_0 {\mid \epsilon \mid}^{-\nu}, \\
\tau = \tau_0 {\mid \epsilon \mid}^{-\mu}, 
\end {eqnarray}

\noindent where $\epsilon$ characterizes the proximity to the critical temperature,
\begin {equation}
\label{eq-epsilon}
\epsilon = \frac {T_c-T} {T_c}, 
\end {equation}

\noindent and $\mu$ and $\nu$ are critical exponents appropriate for the theory under consideration.  
The quench is assumed to occur linearly in time

\begin {equation}
\epsilon=\frac {t} {t_Q}
\end {equation}

\noindent so that for $t<0$, the temperature of the heat bath is above the critical temperature
and the critical temperature is reached at $t=0$. 
The divergence of the equilibrium relaxation time as the heat bath approaches the critical temperature
is known as critical slowing down.
Critical slowing down results from the finite speed of propagation of perturbations of the order 
parameter.  As the correlation length diverges, small 
perturbations of the order parameter ({\it e.g.,} lowering of the temperature) take longer to propagate over correlated 
regions and therefore it takes longer to reach equilibrium.  As the critical temperature is approached from above there comes a time 
$\mid t* \mid $ during the quench when the time remaining before the 
transition equals the equilibrium relaxation time

\begin {equation}
 \mid t* \mid=\tau(t*).
\end {equation}

\noindent Beyond this point the correlation length can 
no longer adjust fast enough to follow the changing temperature of the bath.  At time {\it t*} the dynamics of the
correlation length ``freezes".   The correlation 
length remains frozen until a time {\it t*} after the critical temperature is reached.  
In Zurek's proposal the 
correlation length at the ``freeze-out'' time {\it t*} sets the characteristic length scale
for the initial defect network.  
Solving for the value of the correlation length at {\it t*}, the frozen correlation length, 
and therefore the initial defect density, scale with the quench rate as

\begin {equation}
\xi (t*) \sim {\tau_Q}^{\frac {\nu} {1+\mu}}.
\end {equation}

\noindent Aspects of this scenario have been verified using 1-dimensional and 2-dimensional simulations of phenomenological
time-dependent Landau-Ginzburg equations \cite {ZurLag} \cite{ZurYat}.  Preliminary experimental efforts to test the Zurek 
prediction in condensed matter systems are inconclusive, lacking reliable error estimates and a mechanism to vary the quench time scale \cite {DefExp}.
    
The Kibble prediction for the initial length scale of the defect 
distribution is inconsistent with experiment, yet the alternative  
``freeze-out'' proposal, raises many conceptual questions. The ``freeze-out'' scenario of Zurek relies on 
phenomenological ideas derived from experience with classical 
Landau-Ginzburg systems.  It is not clear how much of this picture, if any, is applicable
to {\it quantum} fields which undergo phase transitions in the early Universe.
Quantum fields bring new conceptual issues. In particular, consideration of 
decoherence is necessary to understand when and how the quantum system at the late stages of the transition 
can be described by an ensemble of classical defect configurations.  
Even in the classical context, the use of {\it equilibrium} critical scaling
in a fully dynamical setting is an approximation.  Critical dynamics is generally much richer and less universal than
equilibrium critical behavior \cite{HohHal}.  For a system coupled to a rapidly changing environment, the correlation length evolves through the
dynamical equations of motion of the field.  Worse still, the relaxation time is not always well-defined. 
In addition, a realistic bath consists of interactions between the system and other fields in the Universe.  Physical interactions 
between the bath and the system produce interesting and nonequilibrium behavior 
such as noise and dissipation \cite{Banff}.  The reduction of the complicated
bath-system interaction to one parameter, the quench time scale, and a prescribed linear time-dependence in the effective 
mass of the time-dependent Landau Ginzburg equation neglects these potentially important processes.

\subsection{Foundational Issues}

Despite significant effort, the physical mechanisms important for the 
formation of topological defects in the early Universe and  
condensed matter systems are not well understood. It is useful to explore three interwoven and 
complementary themes that are relevant to this problem.  These are (i) equilibrium vs. 
nonequilibrium, (ii) microscopic vs. macroscopic, and (iii) quantum vs. classical.

\subsubsection {Equilibrium vs. Dynamical}

The Zurek and Kibble mechanisms both rely on aspects of equilibrium physics. 
However, as emphasized by Zurek, the formation of topological defects is a 
nonequilibrium process. If aspects of the ``freeze-out'' scenario are correct then it must be viewed as an approximation
to a more fundamental and as yet unexplored understanding. The initial density of topological defects formed in 
{\it far-from-equilibrium} phase transitions may be determined by very different mechanisms.

\subsubsection {Microscopic vs. Macroscopic}

Topological defects appear as configurations of the classical order 
parameter field in phenomenological Landau-Ginzburg theories and in the classical limit of symmetry-broken
quantum field theories.  A first-principles approach to the formation of topological defects 
in a nonequilibrium quantum field phase transition requires an understanding of the microscopic origin of 
macroscopic dynamical critical behavior like critical slowing down.  Macroscopic domains of correlated vacuum 
must be identified from the microscopic quantum field.  Thermalization occurs through
microphysical couplings between the quantum field composing the system and other fields composing an environment and
must be understood to produce a physical bath-system interaction.

\subsubsection {Quantum vs. Classical}

Phase transitions occur in quantum field theory and in condensed matter systems. In a second-order phase transition, the 
correlation length diverges at the critical temperature and the dynamics is dominated by low frequency modes. In equilibrium,
modes with $\omega <<T$ have essentially a Maxwell-Boltzmann distribution and it might be expected that second-order phase
transitions are dominated by classical behavior. In a dynamical setting an exact equilibrium distribution does not exist
and a necessary indication of classical behavior is the emergence of 
a positive definite probability distribution function from the density matrix of the quantum field \cite{CoopHab}.   In quantum critical systems 
decoherence may modify critical scaling exponents and change the coarsening of domains from that of classical systems.  Even the concepts 
used to describe phase transitions in condensed matter may not apply in the quantum context.  For example, the definition of a 
topological defect in a quantum field theory with arbitrary quantum state is unknown.

\vspace{5mm}

\noindent Phase transitions and topological defects occur in both classical and 
quantum systems, under equilibrium and nonequilibrium conditions and are depicted by  
phenomenological and microscopic theories. They therefore provide a theoretical vantage point 
from which to further elucidate the connections 
between these complementary and disparate
concepts.  The formation of topological defects is an arena in which these general yet basic issues 
of physics may be studied in a concrete and useful manner.

\subsection{Quantum Field Dynamics and the Formation of Topological Defects in the Early Universe}

A complete understanding of the physical issues involved 
in the formation of topological defects in a 
second-order phase transition in the early Universe requires a
first-principle approach to the nonequilibrium dynamics of quantum fields, 
a realistic treatment of the interaction of
the quantum field both with gravity and with other fields 
that constitute an environment, and the 
identification of classical defect configurations from the quantum field system.

This work will concentrate on the dynamics of the phase transition important for defect formation. 
While symmetry restoration in finite-temperature quantum field theory has been known 
since the early work of
Kirzhnitz and Linde \cite{KirLin}, most previous efforts have
focused on the equilibrium aspects of the transition.  
Techniques such as the 
finite-temperature effective potential \cite{DolJak} \cite{Weinberg} and the renormalization group \cite{RG} have been developed to deduce 
equilibrium critical properties such as the order of the phase transition and its
critical temperature.  However, equilibrium techniques 
are inadequate to study the {\it dynamics} of the phase transition.  The use of equations of 
motion for the mean field derived from the finite-temperature
effective potential was criticized in \cite{MazUnrWal}.   In general the use of equations generated 
from the finite-temperature effective potential in a dynamical
setting results in unphysical solutions.   Although equilibrium techniques
are clearly inappropriate, solving the full equations of motion
for an interacting field theory is generally impossible, even numerically.  
This difficulty is partially overcome by the development of approximation schemes which allow for
the evolution of a restricted set of correlation functions of 
the quantum field theory \cite {CalHu87} \cite{CalHu88}.  These methods have been applied to a number 
of dynamical problems in quantum field theory  
including post-inflationary reheating \cite{RamHu} (and references therein) and spinodal decomposition \cite {Cal} \cite {BoySpin} \cite{CoopHab}.

The problem of defect formation in a nonequilibrium second-order phase transition of a quantum field has 
also recently received attention \cite {GilRiv} \cite{KarRiv} \cite{BowMom}.  The results 
are promising but the studies are incomplete.  In these previous approaches the phase transition is incorporated through 
an {\it ad hoc} time dependence of the effective mass of a {\it free} field theory:  an instability in the theory is
induced when the mass becomes tachyonic.  The use of a prescribed time dependence of the effective mass, 
while providing a convenient analytic model, lacks 
physical justification.  The neglect of interactions confines the applicability
of these approaches to very early times before the field amplitude grows substantially. They are therefore 
unable to account for the back reaction which is necessary to  stabilize domain growth and shut off the spinodal 
instabilities of the phase transition. In addition, defects formed during
the linear stages of the phase transition are transient and not likely to survive to late times.

\subsection{Outline}

In this paper we analyze the formation of correlated domains of true vacuum in the 
time evolution of a quantum scalar field during a second-order phase transition
initiated by cooling of a radiation-dominated FRW Universe.  
In Section II we discuss the formalism used to follow the dynamics of a quantum field through the phase transition. 
We also present the model, a derivation of the equations of motion for the two-point function of the theory, 
and discuss renormalization, initial conditions and numerical parameters used in the numerical simulation.
We then provide a dynamical description of the phase transition using results of the numerical simulations.  
Section III begins the discussion of domains and presents the argument that
domains are determined by a peak in the Fourier space structure function $k^2G(k,t)$.  Section IV discusses the power 
law scaling of the size of domains with the quench rate.  The equilibrium critical exponents of the correlation length and
relaxation time are calculated in both the underdamped and overdamped cases and the power law scaling of the 
defect density with the quench rate predicted by the
``freeze-out'' proposal is shown to be in good agreement with the numerical simulations.  An analytical model valid for slow quenches and near the 
onset of the instability is introduced and the power law exponent in the analytic model is found to be the same as the numerical simulations. 
Section V provides a summary and discussion of 
these results and presents possible directions for further study.  
  
\section {Phase Transition Dynamics}

It is common to model critical dynamics with a 
classical, phenomenological, time-dependent Landau-Ginzburg equation 
for an order parameter $\Psi$,

\begin{equation}
\partial_t \Psi(\vec{x},t) = -\Gamma \frac {\delta F} {\delta \Psi } +\xi,
\end{equation}

\noindent where $F[\Psi]$ is a phenomenological free energy density for the order parameter, $\Gamma$ is a phenomenological dissipative coefficient
and $\xi$ is a stochastic term
incorporating thermal fluctuations of the environment \cite{Gold}. 

Even if the order parameter is of quantum origin, as {\it e.g.} in the phase of the 
wavefunction for liquid Helium-4, the Landau-Ginzburg equation is rarely 
derived logically from the underlying quantum dynamics of the system.  
In condensed matter systems, to compensate for insufficient microscopic information, great care with physical intuition goes 
into choosing the order parameter and its equation of motion.  In 
experimentally inaccessible environments, such as the early Universe, it is
not {\it a priori} obvious what the order parameter, or its dynamics, should be.  
In situations where phenomenological approaches are inadequate, it is 
necessary to work with the fundamental quantum dynamics of the fields. 

A first-principles approach to the quantum dynamics of phase 
transitions avoids {\it ad hoc} assumptions about the dynamics of the correlation length and
the effect of the quench.  The system simply evolves under the true 
microscopic equations of motion.  We can therefore explore many 
details of critical dynamics that are inaccessible in phenomenological theories.     

Solving the exact dynamics of interacting quantum fields expressed, for example, 
through the Heisenberg equations of motion for the field operator is a 
very complicated problem.  Instead, we seek a truncation of the full degrees of 
freedom that allows for an accurate modeling of the dynamics of the phase transition 
over the time scales of interest.

Consider a classical $\Phi^4$ scalar field theory with $m^2>0$ and self-coupling $\lambda$.
The potential $V[\Phi]$ for the field is

\begin {equation}
V[\Phi]= -\frac {1} {2} m^2  \Phi^2+\frac {\lambda} {4} \Phi^4.
\end {equation}

\noindent The degenerate true minima of this potential are

\begin {equation}
\Phi_{min} =  \pm \frac {m} {\sqrt{\lambda}},
\end {equation}

\noindent which are nonperturbatively large in the coupling constant.  To follow the field as it evolves from the unstable vacuum 
$\Phi=0$ to the stable one $\Phi= \pm \frac {m} {\lambda}$ requires an approach which is both non-perturbative in the field amplitude
and fully dynamical.

A useful formalism that is both nonperturbative and dynamical is the 
two-particle irreducible closed-time-path (2PI-CTP) or in-in effective action \cite {CalHu87} \cite{CJT}. 
The 2PI-CTP effective action generates real and causal equations of motion for the mean field $\langle \Phi(\vec{x},t) 
\rangle$ and two-point correlation functions $\langle \Phi(\vec{x},t) \Phi(\vec{y},t') \rangle$ of the quantum field 
theory.  A diagrammatic expansion of the 2PI effective action, to arbitrary loop order is given in \cite{CJT}.  
Calculated to all orders in a loop expansion, the 2PI-CTP effective action and the equations of motion derived 
from it contain all information of the original quantum field theory. However, any practical computation requires 
inclusion of terms only to some finite loop order which
constitutes an approximation \cite{CalHu88} \cite{Winn}.  In this paper, we neglect three-loop and higher graphs.  The truncation of the 2PI-CTP 
effective action to two-loop order is equivalent to the time dependent Hartree-Fock approximation \cite {CJT}.

The equations of motion for the mean field and the two-point function derived from the 2PI-CTP effective action respect the 
$\Phi \rightarrow -\Phi$ symmetry of the classical action.  Since the field starts in a symmetry-restored state 
above the critical point where

\begin{equation}
{\langle \Phi \rangle}_{initial} =0,
\end{equation}

\noindent the mean field remains identically zero throughout the phase transition. The dynamics of the phase transition
unfolds through the dynamics of the two-point function.  

The choice to limit attention to the 
two-point correlation functions in the Hartree-Fock approximation limits the time 
scale over which the evolution is physically reliable.  Higher order correlations are suppressed in weakly coupled field theories in
equilibrium.  However, second-order phase transitions are characterized by spinodal instabilities that cause 
correlations to grow.  When the correlations grow large enough that the field is sampling the stable vacua, dissipative   
processes are expected to become important and the Hartree-Fock approximation will break down \cite{BoySpin}.

\subsection{The Model}           
We consider a scalar field in an FRW spacetime \cite{BoyFRW1} \cite{BoyFRW2}. 
The field has the symmetry-breaking classical action

\begin {equation}
S=\int  d^4x \sqrt{-g}(\partial_{\mu} \Phi \partial^{\mu} \Phi +m^2 \Phi^2 -\frac {\lambda} {4!} \Phi^4)
\end {equation}

\noindent where $g$ is the determinant of the metric of the classical 
background  spacetime.  We assume 
that the stress-energy tensor is dominated by other radiation fields present 
in the early Universe.  These fields maintain the overall 
homogeneity and isotropy of the Universe. Small deviations in 
homogeneity and isotropy that may eventually be responsible for the fluctuations observed in the
cosmic microwave background radiation are produced in our model by the topological 
defects of the system and appear only at the end of the phase transition.
We therefore work in the semiclassical test-field approximation (ignoring 
back reaction of the $\Phi$ field on the spacetime), and we assume 
that the scale factor has the time-dependence of a homogeneous and isotropic, spatially flat,
radiation-dominated Universe,

\begin {equation}
\label {eq-scale}
a(t)= \left[ \frac {t+\tau} {\tau} \right ] ^{\frac {1} {2}}.
\end {equation}

\noindent The expansion of the Universe and the resulting redshifting of the modes act here as a physical
quench allowing the dynamics of the phase transition to unfold naturally.  This is in distinction to work which uses an 
instantaneous change in the sign of the square of the mass \cite {BoySpin} \cite{GilRiv}.

In the Heisenberg representation the field operator $\Phi_H(\vec{x},t)$ can be written as

\begin{equation}
\Phi_H(\vec{x},t)=\int d^3 \vec{k} \left [
e^{i\vec{k}\cdot\vec{x}}f_k(t)a_{\vec{k}} + e^{-i\vec{k}\cdot\vec{x}}f^*_k(t)a^{\dag}_{\vec{ k}}\right ]
\end{equation}
\noindent where the $f_k$ are the complex quantum modes of the field.

In a FRW Universe, the two-loop 2PI equation of motion in 
comoving cosmological time for the mode function with comoving momentum k is

\begin {equation}
\label {eq-mode}
\left (\frac {d^2} {dt^2} + 3\frac {\dot{a}(t)} {a(t)} \frac {d} {dt} + \frac {k^2} {a^2(t)}-m^2
+ \frac {\lambda} {2} G(t,t)\right )f_k(t) = 0,
\end {equation}
 
\noindent where

\begin {equation}
G(t,t) = \int \frac {d^3k} {(2\pi)^3} f_k(t)f^*_k(t)\sigma_k(\beta)
\end {equation}

\noindent is the equal-time limit of the two-point correlation function and

\begin {equation}
\sigma_k(\beta)=\coth(\frac {\beta}{2}w_k(0))
\end {equation}

\noindent is a constant factor incorporating thermal initial conditions with temperature $T=1/\beta$.  The
effective mass of the system is 
\begin{equation}
m^2_{eff}(t)=-m^2+\frac {\lambda} {2} G(t,t)
\end{equation}
\noindent and the initial frequency is

\begin {equation}
w^2_k(0)=k^2+m^2_{eff}(0),
\end {equation}

\noindent where $m^2_{eff}(0)$ is the initial finite-temperature effective 
mass in the Hartree-Fock approximation. Since we work in a radiation-dominated FRW universe, the scalar curvature, $R$, is zero and
the conformal coupling constant $\xi$ may be ignored.

\subsubsection{Renormalization}

The equal-time limit of the two-point function is divergent and must be 
regularized.  A simple regularization method, amenable to a numerical simulation,
is to implement an ultraviolet cutoff in physical spatial momentum.  
A suitably regularized expression for the two-point function 
must be independent of this cutoff. Our renormalization scheme follows that of \cite{BoyFRW1} \cite{BoyFRW2}. 

It is a general feature of 
quantum field theory that the bare quantities appearing in the classical 
Lagrangian are not observable but are ``dressed'' by interactions.  In the Hartree approximation, the effect of interactions is encoded in
the self-consistent effective mass. Therefore, there are no counterterms to absorb divergences and the equation for the mode 
functions, Eq. ({\ref{eq-mode}), must
be finite \cite{Root}. We therefore fix the renormalization scheme with the condition

\begin {equation}
-m_B^2+\frac {\lambda_B} {2} G_B(t,t)=-m_R^2+\frac {\lambda_R} {2} G_S(t,t).
\end {equation}

\noindent The cutoff dependence of the bare variance is obtained  by
considering a WKB-type solution to the mode function equation. 
Identifying the second order adiabatic mode functions from the 
WKB solution of the mode function
equation, the subtracted two-point function is 

\begin{equation}
G_S(t,t)=\frac {1} {2\pi^2} \int_{0}^{\Lambda} k^2 dk \left ( f_k f^{*}_k \sigma_k (\beta)
 -\frac {1} {ka^2(t)} + \frac {\theta(k-\kappa)} {4k^3} m^2_{eff} \right )
\end{equation}

\noindent In combination with the renormalization condition, this 
subtraction can be implemented by
a shift in the bare parameters of the theory

\begin{eqnarray}
m_b^2 + \frac{\lambda_b}{16\pi^2}\frac{\Lambda^2}{a^2(t)} = 
m_r^2 \left[1+\frac{\lambda_b}{16\pi^2}\ln(\Lambda/\kappa)\right], \\
\lambda_b = \frac{\lambda_r}{1-\frac {\lambda_r} {16\pi^2} \ln(\Lambda/\kappa)}.
\end{eqnarray}

\noindent The shift is time independent as long as the cutoff $\Lambda$ 
and the renormalization scale $\kappa$ are implemented in terms of the physical
momentum \cite{BoyFRW1}

\begin {eqnarray}
\Lambda=\Lambda_0 a(t) \\
\kappa=\kappa_0 a(t)
\end {eqnarray}
 
\noindent where $\Lambda$ and $\kappa$ are comoving and $\Lambda_0$ and
$\kappa_0$ are physical quantities.  The renormalized mode function equation is

\begin{equation}
\label {eq-rmode}
\left (\frac {d^2} {dt^2} + 3\frac {\dot{a}(t)} {a(t)} \frac {d} {dt} + \frac {k^2} {a^2(t)}-m^2_r
+ \frac {\lambda_r} {2} G_S(t,t)\right )f_k(t) = 0.
\end{equation}

\noindent In following sections we will drop the renormalization subscripts for clarity and it is to be 
understood that we are working with renormalized quantities.

\subsubsection{Initial Conditions}

The quantum field is assumed to be initially in a state of thermal equilibrium. 
In this model with a tachyonic tree-level mass, symmetry is restored by finite 
temperature corrections.  The initial effective mass $m^2_{eff}$ is the solution of the 
(renormalized) equation

\begin{equation}
\label {eq-minit}
m_{eff}^2=-m^2+\frac{\lambda} {4\pi^2} \int_{0}^{\Lambda} k^2 dk \left (\frac {\coth{\frac{\beta \sqrt{k^2+m_{eff}^2}} {2}}}
{2\sqrt{k^2+m_{eff}^2}} -\frac {1} {2k} + \theta(k-\kappa)\frac{ m^2_{eff}} {4k^3} \right ).
\end{equation}

\noindent In the high temperature and small-$\lambda$ limit this yields

\begin{equation}
\label {eq-highTm}
m_{eff}^2=-m^2+\frac {\lambda T^2} {24},
\end{equation}
\noindent which is a result familiar in finite-temperature field theory \cite{KirLin}

In an expanding FRW Universe, exact thermal 
equilibrium will only persist for conformally invariant fields.  
If the expansion rate is small relative to internal collisional processes of the field 
then there is an approximate notion of equilibrium \cite{equil}.  
This is evidenced by transforming to conformal time, $\eta$, defined by 

\begin{equation}
dt=a(\eta)d\eta,
\end{equation}
                                                                                               
\noindent and performing a mode redefinition
\begin{equation}
\tilde{f}_k(\eta)= f_k(\eta)a(\eta)    
\end {equation}

\noindent The conformal mode function equation is now

\begin {equation}
\label {eq-conmode}
\left (\frac {d^2} {d\eta^2} + k^2 + a^2(\eta)(m^2+\frac {\lambda} {2} G(\eta,\eta))\right )\tilde{f}_k(\eta)=0.
\end {equation}

\noindent If the Universe is slowly expanding,  a WKB-type solution 
is appropriate and a low-adiabaticity truncation of the instantaneous
WKB frequency is sufficient. The zeroth-adiabatic order solution to
equation (\ref {eq-conmode}) is given by 

\begin{eqnarray}
\tilde{f}(\eta)=\frac {1} {\sqrt{2w_k}} e^{-i\int^{\eta}w_kd\eta'}, \\
w_k^2=k^2+a(\eta)^2m_{eff}^2.
\end {eqnarray}

\noindent This leads to the following initial conditions for the 
mode functions in cosmic time t
\begin{eqnarray}
f_k(0)=\frac {1} {a(0)\sqrt{2w_k(0)}}, \\
\dot {f_k}(0) = \left(-\frac {\dot{a(0)}} {a(0)} -iw_k(0)\right)f_k(0).  
\end{eqnarray}

\noindent Slow expansion assumes that the 
natural frequency of the kth mode is faster than 
the expansion rate of the Universe,

\begin {equation}
w_k(0) \gg \frac {1} {2\tau}.
\end {equation}

\noindent The adiabatic equilibrium approximation would fail for the 
lowest k modes.  However 
with high temperature initial conditions the low k modes are not a 
dominant part of the spectrum. 

\subsubsection{Numerical Parameters}

The model described by Eq.(\ref{eq-mode}) and (\ref{eq-scale}) is characterized by 6 parameters,
\begin {equation}
(m,\: \tau, \: \lambda, \: T, \: \Lambda, \: \kappa).
\end {equation}
 
\noindent We choose units of energy in which 

\begin {equation}
m^2=1.0.
\end {equation}

\noindent The initial rate of expansion or the initial Hubble 
constant is controlled 
by $\tau$,

\begin {equation}
H(0)=\frac {\dot{a}(0)} {a(0)}=\frac {1} {2\tau}.
\end {equation}

\noindent In the simulations we conducted the range of the $\tau$ parameter was

\begin {equation}
0.01\leq \tau \leq 100
\end {equation}

\noindent and the self-coupling is strong,
\begin {equation}
\lambda=0.1.
\end {equation}

\noindent The initial temperature $T$ is chosen
so that the value of the initial effective mass $m^2_{eff}(0)$ is of order unity.  More specifically
we choose

\begin {eqnarray}
m^2_{eff}(0)=-1.0+\frac {\lambda} {24} T^2 = .607 \\
T=20.0
\end {eqnarray}

\noindent  The values of the initial effective mass $m^2_{eff}(0)$, the coupling $\lambda$ and expansion rate $\tau$, 
were chosen so that the simulations of the phase transition completed on numerically accessible timescales. Due to the 
renormalization scheme, both the comoving cutoff $\Lambda$ and the comoving renormalization scale $\kappa$ 
increase with the scale factor.  The initial 
value 

\begin {equation}
\Lambda_0 = 340
\end {equation}

\noindent was chosen as the lowest value that maintained cutoff 
independence of the initial effective mass.  Insensitivity to the cut-off in the dynamical simulations
was verified by doubling $\Lambda_0$ and observing no change in the output
plots of the time-dependent effective mass.  The initial value of the 
renormalization scale 

\begin {equation}
\kappa_0=1.0
\end {equation}

\noindent was chosen so that the renormalization scale was always above 
the maximum momentum of the unstable modes.  The coupled, nonlinear system of mode function equations with the 
given initial conditions was solved numerically using an adaptive stepsize,
fifth-order Runge-Kutta code.  Mode integrals were performed using a simple simpson rule 
with a uniform momentum binning

\begin {eqnarray}
k=nk_{bin}, \\
k_{bin}=\frac {2\pi} {L_0},
\end {eqnarray}

\noindent where $L_0=100.0$ is the effective size of the system and n is the total number of modes.  
Insensitivity to the momentum binning was verified by reducing $k_{bin}$ and 
observing no change in output. The number of
modes $n$ varied from $10^4$ to $10^5$. Run times varied from hours to 
days on a DEC 500 MHz workstation which corresponds to dynamical time scales of $t=2$ to $t=100$.      

\subsubsection {Analysis}

The results of a typical simulation are shown in Figs. 1 and 2.  In Fig. 1 
the renormalized effective mass is shown as a function of 
cosmological time for quench parameter $\tau=1.0$.  The phase 
transition
begins when $M_{eff}^2$ first becomes negative. When $M_{eff}^2$ is negative,
modes with physical momentum $\frac {k} {a} \leq M_{eff}$ 
have imaginary frequencies and begin to grow.  This indicates the onset of 
the spinodal instability which is characteristic of a second-order phase 
transition. In the early stages of the phase transition, the evolution of the 
effective mass is dominated by the redshifting of stable modes 
and the effective mass decreases. As the phase 
transition proceeds, more modes redshift into the unstable momentum band and 
the amplitude 
of unstable modes continues to grow.  Eventually, the redshift of stable 
modes balances the growth of unstable modes and the effective 
mass increases.  As the effective mass passes through zero from below, 
the field reaches the spinodal point. Beyond the spinodal point 
all modes are stable. As the effective mass continues to grow, 
non-linear thermal and other collisional processes are expected to be
important and the Hartree-Fock approximation of the dynamics of the two-point
correlation function breaks down (see for example \cite{Cal}).

In Fig. 2 we show the Fourier space structure factor 

\begin{equation}
S(k,t) \equiv k^2G_k(t,t)
\end{equation} 

\noindent at various times during the phase transition.  As argued in the
next section, defects and domains can be identified in the low-k structure of 
$S(k,t)$. As the phase transition begins, $S(k,t)$ develops a peak at 
low k. As the phase transition proceeds, this peak 
grows in amplitude and redshifts until at late times it completely dominates the 
infrared portion of the spectrum.

\section {Defect Density}

We have argued that the dynamics of a phase transition of a quantum field 
may be approximated by the dynamics of the two-point function 
at least for times less than the spinodal time. Using the two-loop 2PI 
equations of motion we obtained a numerical solution for the evolution 
of the two-point function. To observe the formation of correlated domains 
and topological defects it is necessary to identify these structures 
from the form of the two-point function.

The identification of topological defects from the
underlying quantum dynamics of the two-point function is a complicated
problem.  Intuitively, the existence of well-defined 
correlated domains of true vacuum at the 
completion of the phase transition is similar to the domains that form 
in a condensed matter system like a ferromagnet.  However, our system is described by a 
{\it quantum} field theory and the existence of a classical configuration of 
domains requires a quantum-to-classical transition, of which decoherence is an essential condition. 
The extraction of a positive-definite probability distribution, $P[\Phi]$, from the density matrix
requires a quantitative treatment of the quantum-to-classical transition.  Classical defect solutions would then appear in
field configurations drawn from $P[\Phi]$ \cite{CoopHab}.   

In this research our focus is {\it not} on the extraction of classical defects from the quantum system but instead on the long wavelength
modes that determine the size of correlated domains and, therefore, the average defect separation.  The long wavelength modes interact
with an environment of short wavelength and thermal fluctuations that destroys quantum coherence among the long 
wavelength modes and results in a finite correlation length for the system \cite{HuPazZhang}.  While quantum fluctuations are indeed important to
the description of phenomena within the scale of one domain, beyond this scale we may hope to treat the modes as classical.  With
this abbreviated estimate of decoherence we show that the existence and size of correlated domains is indicated by 
an infrared peak in the power spectrum of the equal-time momentum-space two-point function. 

For a free field theory quenched into the unstable region by an 
instantaneous change in the sign of the square of the mass at $t=t_0$ it is 
possible to obtain analytic expressions for the equal-time momentum 
space two-point function \cite {BoySpin}.  After the quench, the momentum space structure function $k^2G(k,t)$ 
has a strong maximum at the value

\begin {equation}
k_{max} \sim \sqrt{\frac {m_f} {2t}}.
\end {equation}

\noindent The real space Fourier transform of this function 
(normalized to unity at $\left | \vec{r}-\vec{r'} \right |=0$) is

\begin {equation}
G(\left | \vec{r} -\vec{r'} \right |,t) \sim e^{-\frac {m_f(\left |
\vec{r}-\vec{r'} \right |)^2} {8t}}
\end {equation}

\noindent This form of the equal-time correlation function is common in condensed matter systems \cite {Bray}. 
From the equal-time real space correlation
function we can identify the correlation length

\begin{equation}
\xi(t) \sim \sqrt{ \frac {8t} {m_f}} \sim \frac {1} {k_{max}}
\end{equation}

\noindent In analogy with the free field theory, we assume the 
domain size is proportional to the size of the maximum of the momentum space structure function.

Even when the density matrix of the quantum field has decohered into a positive-definite
probability distribution for an ensemble of classical field configurations, the identification of 
topological defect structures is a 
complicated task.  In a classical field configuration and when 
defects are well-formed (so that the width of the defect is much smaller 
than the typical defect spacing), it is possible to identify zeros of the 
classical field configurations with topological defects.  In a model 
with a global $O(n)$ symmetry in $n$ spatial dimensions, 
and when the field probability distribution is Gaussian, a formula for the 
ensemble average density of field zeros  
was given by Halperin \cite {Hal} 
and derived explicitly by Liu and  Mazenko \cite {LiuMaz}:

\begin{eqnarray}
\label {eq-density}
\rho(t)= C_n\left (\frac {G''(0,t)} {G(0,t)}\right )^{n/2} \\
C_1=\frac{1}{\pi},\: C_2=\frac{1}{2\pi},\: C_3=\frac {1}{\pi^2}\nonumber
\end {eqnarray}

\noindent For a single scalar field in 3 spatial dimensions 
Eq. ( \ref{eq-density}) with $n=1$ is
valid as the {\it ensemble-averaged density of zeros along a one-dimensional 
section of the field}.  The validity of the Gaussian approximation is further discussed
in \cite{CalIba}.

A zero of the classical field configuration does not 
uniquely identify a topological defect.  Thermal 
fluctuations give a large number of zeros of the field configuration 
on small scales \cite{AlexHabKov}.   The zeros of the field 
configuration which are not associated with 
different defects lead to an ultraviolet momentum divergence
in the expression for the zero density, Eq. (\ref{eq-density}).  To make 
physical sense of the zero formula, a coarse graining of the 
field configuration is needed.  The spacing 
of defects and the size of correlated domains is determined by long 
wavelength excitations and if zeros of the 
field configuration are to accurately count defects, it is 
necessary to apply a coarse-graining that removes the short-wavelength 
excitations caused by quantum and thermal fluctuations.  To 
effect this coarse-graining in Eq.(\ref{eq-density}) we impose a 
spatial momentum cut-off at the upper edge of the unstable momentum band. Only the 
unstable modes will contribute to the domain size. 

In principle it is possible that the length scales set by $k_{max}(t)$ and 
by $\rho(t)$ are different. When the structure factor $k^2G(k,t)$ is very strongly peaked about $k_{max}$ 
we can approximate the unstable portion of the spectrum as
\begin {equation}
k^2G(k,t) \sim \delta (k-k_{max})
\end {equation}

\noindent the zero density for a single scalar field in 3 dimensions is then
\begin {equation}
\rho(t) = \frac {1} {\pi} \left( \frac {\int k^4 G(k,t)} {\int k^2 G(k,t)}\right)^{3/2} \sim k_{max}(t)^3.    
\end {equation}

\noindent For the parameters used in the simulation
and at late times, the scales set by the zero density and by the domain size differ only by a 
numerical factor of order unity.  Using the linear model introduced in Sec. V,  evidence that the growth of domains
is characterized by one scale is shown in Fig. 5.  Additional evidence for a ``one-scale'' model was given in 
\cite{AntBet} in the context of the dynamics of a classical field phase transition in $1+1$ dimensions.

For very weak coupling not considered here and when a peak is no 
longer evident in $S(k,t)$,
the one-scale approximation  will fail and it is possible in principle 
for the defect density to depart significantly from
one per correlation volume \cite{KarRiv}.

\section {Critical Scaling}

The evolution of the two-point function and the extraction of 
correlated domains and topological defects allows a first-principles 
analysis of the mechanisms important for topological
defect formation.  In light of the ``freeze-out'' scenario it is 
interesting to compare the initial size of correlated domains for 
scale factors $a_\tau(t)$ with different
expansion rates $\tau$.  To quantify the dependence of the initial size of 
correlated domains on the quench rate $\tau$ of the phase transition, we 
compare the size of domains for different values of the parameter $\tau$. As 
discussed above, the average size of a domain is proportional to the
maximum in the infrared portion of $S(k,t)$. 
We compare the maximum $k_{max}$ for different 
values of $\tau$ and at two distinct sets of times during the phase transition. 
The first set of domains is measured when the square of the effective 
mass reaches a minimum value.
This provides an early-time measure of the size of domains.  The results are 
shown in Fig. \ref{fig-eslowpowerlaw}.  The second set of domains is measured when the square of
the effective mass reaches a local maximum value after the phase transition.  
This provides a late-time measure 
of the size of domains.  The results are shown in Fig. \ref{fig-lslowpowerlaw} and Fig \ref{fig-lfastpowerlaw}.  
For slow quench rates ($\tau \geq 1.0$), at both early and late times, the dependence of $k_{max}$ on the 
quench rate $\tau$ is well approximated by a power law,

\begin{equation}
k_{max}(\tau) \sim \tau^{-0.35}.
\end{equation} 

\noindent  For fast quench rates ($\tau<1.0$), the dependence of $k_{max}$ on the quench rate $\tau$ is well approximated by a power law,
\begin{equation}
k_{max}(\tau) \sim \tau^{-0.28}.
\end{equation}
 
\noindent The origin of these exponents and the difference between slow and fast quenches is discussed in the following
section.

\subsection {The Freeze-Out Scenario}

In Zurek's scenario the ``frozen'' correlation length and therefore 
the initial size of correlated domains scale as a power law of the 
quench time $\tau$
as

\begin {equation}
\label{eq-powerlaw}
\xi_{freeze}  \sim {\tau}^{\frac {\nu} {1+\mu}}
\end {equation}

\noindent where $\mu$ and $\nu$ are the equilibrium critical exponents for 
the correlation length and relaxation time respectively. The correlation length and relaxation time are identified as,
respectively, the length and time scales that characterize the equilibrium
behavior of the propagator near the critical temperature \cite{Lub}.
Under the scaling hypothesis, the equal-time propagator is
written

\begin {equation}
G_k=k^{-(2-\eta)}F[h(\epsilon)k].
\end {equation}

\noindent Here $\epsilon$ is the reduced temperature, Eq. (\ref{eq-epsilon}), $\eta$ is a critical exponent (not to be confused
with conformal time) and $F$ is a dimensionless function.  The finite-temperature equilibrium correlation length is

\begin {equation}
\xi(\epsilon) = h(\epsilon) \sim \epsilon^{-\nu}.
\end {equation}

\noindent Similarly the two-time propagator is
\begin {equation}
G_{\omega, k} = \omega^\alpha F[h'(\epsilon) \omega,h(\epsilon)k]
\end {equation}

\noindent so that the relaxation time

\begin {equation}
\tau \sim  h' \sim \epsilon^{-\mu}.
\end {equation}

\noindent In the Hartree-Fock approximation the equilibrium propagator is
\begin {equation}
G_{\omega, k} \sim \frac {1} {\omega^2-k^2-m^2_{eff}} \coth(\frac {\sqrt{k^2+m_{eff}^2}} {2} \beta)
\end {equation}

\noindent where the effective mass is given by the equilibrium value Eq. (\ref{eq-highTm}).  Near the critical point
\begin {equation}
m^2_{eff}(\epsilon)=A\epsilon,
\end {equation}
\noindent where A is a constant independent of temperature. The scaling
behavior of the propagator is
as $\frac {\omega} {\sqrt{\epsilon}}$ and $\frac {k} {\sqrt{\epsilon}}$ and
therefore the critical exponents for the theory are

\begin{equation}
\label{eq-qcritexp}
\mu=\nu=\frac {1} {2}.
\end{equation}

\noindent The critical exponents derived for this {\it quantum} field theory in 4 {\it spacetime} dimensions are the same as 
the critical exponents of those of a {\it classical} Ising model in 4 {\it space} dimensions since, after the Euclidean 
continuation $t \rightarrow i\beta$, they
are in the same universality class.  The finite size $\beta$ of the Euclideanized time dimension can alter the critical
behavior of the quantum field \cite{HuOCon}. At high temperature or near a second-order phase transition 
equilibrium quantum systems in $d$ spacetime dimensions undergo dimensional reduction and behave with the critical exponents of classical systems 
in $d-1$ space dimensions \cite{EnvRen}.  The
dimensional crossover can be parameterized by

\begin {equation}
z= \xi_T T,
\end {equation}

\noindent where $\xi_T$ is the equilibrium correlation length and T is the
temperature.  The effective critical exponents are functions
of $z$ so that $z=0$ corresponds to the quantum case and $z=\infty$
corresponds to the classical case.  Intuitively, 
dimensional crossover occurs in a second-order phase transition when the correlation length $\xi_T$ is much larger than the Euclidean
compactified dimension $\beta=1/T$.  The quantitative effect of dimensional reduction on the critical exponents has been calculated 
in detail in \cite{EnvRen} where plots of $\nu_{eff}(z)$ and $\mu_{eff}(z)$ may be found.  In the extreme dimensionally-reduced case, 
when $z\rightarrow \infty$, the critical exponents change to

\begin{equation}
\label{eq-ccritexp}
\mu=\nu\approx 0.64.
\end {equation}

\noindent However, in our model, the {\it dynamics} of the system and in particular the ``freeze-out'' of the correlation length prevent $z$ 
from becoming too large and the actual critical exponents lie somewhere between the four dimensional case Eq. (\ref{eq-qcritexp}) and the 
three dimensional case Eq. (\ref{eq-ccritexp}). The ``freeze-out'' prediction for the initial scale of the density of topological
defects, Eq. (\ref{eq-powerlaw}), is relatively insensitive to this change in critical exponents and varies from

\begin{equation}
\xi_{freeze} \sim \tau^{0.33}
\end {equation}

\noindent in the four dimensional case to

\begin{equation}
\xi_{freeze} \sim \tau^{0.38}
\end{equation}

\noindent in the three dimensional case. Our observation of the power law exponent of $0.35$ in the numerical simulations for slow quenches, $\tau \geq 1.0$, (Fig. 3 and Fig. 4) is consistent with these cases.  

The calculation of the equilibrium critical exponent for the relaxation time depends on the dynamics of the mode functions for the the low k modes
near the critical point and in particular whether the dynamics is overdamped or underdamped.   Near the critical point, 
$m^2_{eff} \approx 0$, and the mode function equation is approximately

\begin {equation}
\label {eq-critmode}
\left (\frac {d^2} {dt^2} + 3\frac {\dot{a}(t_c)} {a(t_c)} \frac {d} {dt}
+ \frac {k^2} {a^2(t_c)}
\right )f_k(t) = 0.
\end {equation}

\noindent The critical time $t_c$ can be estimated as the time when the equilibrium effective mass, Eq. (\ref{eq-equilmass}),
goes to zero, 

\begin{equation}
t_c=\frac {2} {3} \tau,
\end{equation}

\noindent and the value of the scale factor at $t_c$ is

\begin {equation}
a(t_c)= \left[ \frac {t_c+\tau} {\tau} \right ] ^{\frac {1} {2}} 
\approx 1.3.
\end {equation}

\noindent The mode function equation near the critical point Eq. 
(\ref{eq-critmode}) is that of a damped harmonic oscillator with natural frequency
$\omega_0=\frac {k} {a(t_c)}$ and damping constant $\Gamma$ where

\begin{equation}
\Gamma=3H(t_c) \approx \frac {1} {\tau}.
\end {equation}  

\noindent The modes for which $\omega_0 > \Gamma$ or 

\begin {equation}
\label{eq-dampcond}
k> \frac {1.3}{\tau}
\end {equation} 

\noindent are underdamped.  The range of wavenumbers $k$ for the modes that determinex
the size of correlated domains also depends on $\tau$. Eq.
(\ref{eq-dampcond}) is 
a condition on $\tau$ such that when

\begin{equation}
\tau > \tau_*
\end {equation}

\noindent the modes responsible for domain formation are underdamped and $\tau_*$ is to be
determined.  For the underdamped case we expect  
$k_{max} \sim \tau^{-\frac {1} {3}}$.  Using $k_{max}$ as a representative wavevector of the modes
responsible for domains, Eq. ({\ref{eq-dampcond}) implies the underdamped condition

\begin {equation}
\tau_* \sim 1.
\end {equation}   

\noindent Slow quenches with $\tau \geq 1.0 \:$ are therefore underdamped. Fast quenches are overdamped and the dynamics of the mode functions 
will be dominated by the first time derivative.  In the overdamped case, the critical exponent, $\mu$, now assumes the non-relativistic value

\begin{equation}
\mu_{overdamped}=2\mu_{underdamped}.
\end{equation}      

\noindent Due to the uncertainties of dimensional reduction, in the overdamped case, the initial scale of the density of topological defects ranges
from

\begin{equation}
\xi_{freeze} \sim \tau^{0.25}
\end{equation}

\noindent for the four dimensional critical exponents to

\begin {equation}
\xi_{freeze} \sim \tau^{0.28}
\end {equation}

\noindent in the three dimensional case.  This is in good agreement with the power law exponent of $0.28$ measured for 
overdamped, fast quenches ($\tau <1.0$) shown in Fig. \ref{fig-lfastpowerlaw}.

In both the overdamped and underdamped cases, the prediction of the ``freeze-out'' proposal for the power law exponent of the
scaling of the initial defect density with the quench rate appears to be in excellent agreement with the numerical simulations.

\subsection {Early-Time Linear Approximation}

If the initial size of correlated domains and the initial density of 
topological defects are determined by processes occurring very near 
the critical point, as claimed in the ``freeze-out'' proposal, then the 
power law scaling of domains with the quench parameter $\tau$ observed 
in the simulation of the Hartree-Fock mode function Eq. (\ref{eq-rmode}) 
will also appear in a linear approximation.  A linear amplitude approximation to the 
full mode function equation is valid for slow quenches and only near the onset of the 
spinodal instability, before the unstable modes have grown appreciably and 
back reaction is important. Consider the linear equation

\begin {equation}
\label {eq-linear}
\left[ \frac{d^2}{dt^2}+3\frac{\dot a}a\frac d{dt}+\frac{K^2}{a^2}%
-m^2\right] F_k=0
\end {equation}

\noindent where
\begin {equation}
K^2=k^2+k_0^2,
\end {equation}

\noindent and $k_0$ is related to the initial temperature 
\begin {equation}
k^2_0=\frac {\lambda} {24} T^2_0
\end {equation}

\noindent In conformal time, we have

\begin {equation}
\eta ^2=4\tau \left( t+\tau \right) 
\end {equation}

\noindent with conformal modes, $f_k$, defined by

\begin {equation}
f_k\equiv F_k a.
\end {equation}

\noindent Eq.(\ref {eq-linear}) may be rewritten in conformal time as

\begin {equation}
\left [ \frac {d^2} {d\eta^2} + K^2 -\frac {m^2\eta^2} {4\tau^2} \right ] f_k=0.
\end {equation}

\noindent This equation can be solved exactly in terms of parabolic 
cylinder functions.  However,
a simpler solution is obtained for times near the critical point.
We further approximate it by

\begin {eqnarray}
\left[ \frac{d^2}{d\eta ^2}-\frac{mK}\tau \left( \eta -\eta _k\right)
\right] f_k=0 \\ 
\eta _k\equiv\frac{2K\tau } {m}
\end {eqnarray}

\noindent The properly normalized solution with vacuum boundary conditions
before the instability is

\begin {equation}
\label {eq-amode}
f_k=i\sqrt{\frac{2x}{3\pi }}K_{1/3}\left[ -\frac {2} {3} \sqrt{\frac{mKx^3} {\tau} }
\right] 
\end {equation}

\noindent where

\begin {equation}
x=\eta -\eta _k
\end {equation}

\noindent and $K_{1/3}$ is a modified Bessel function.

We now use the analytic solution Eq. (\ref{eq-amode})
to examine the dependence of the position of the peak in the 
Fourier space structure factor $k^2G(k,t)$ on the quench parameter $\tau$.
The location of this peak $k_{max}(t)$ redshifts throughout 
the phase transition, moving towards lower momentum as the domains coarsen.
In order to compare domains formed in the linear model with those of the 
Hartree evolution we compare the position of the peaks at very early times when a peak
is first identifiable in $S(k,t)$.  Plots of $S(k,t)$ with the analytic modes were made with
{\it Mathematica}. The results are 
shown in Fig. \ref{fig-apowerlaw}.  The scaling of domains with the 
quench parameter $\tau$ with the same power law exponent as seen in the full 
numerical simulation for slow quenches is evident. 

\section {Discussion and Conclusion}

Using the Hartree-Fock approximation to the equations of motion for 
the two-point function of a quantum scalar field undergoing a phase 
transition in a 3+1 dimensional spatially flat, radiation-dominated FRW Universe we have 
shown that the size of correlated domains, measured as the maximum of the
peak of the infrared part of the spectrum of $k^2 G(k,t)$, scales as a power 
of the quench rate $\tau$ as

\begin {equation} 
\xi_{domains} \sim \tau^{0.35},
\end {equation}

\noindent for slow, underdamped quenches ($\tau \geq 1.0$) and

\begin{equation}
\xi_{domains} \sim \tau^{0.28},
\end{equation}

\noindent for fast, overdamped quenches ($\tau<1.0$).  The observed power law scaling of correlated domains is 
quantitatively consistent with the ``freeze-out'' hypothesis.  In both overdamped and underdamped cases, the value of 
the power law exponent extracted from the Hartree-Fock evolution of the two-point 
function is in good agreement with the value calculated in the ``freeze-out''
scenario using the critical scaling exponents for a $\Phi^4$ 
theory in the Hartree-Fock approximation.

To further explore the behavior of the quantum system near the 
critical point we introduced an approximate linear model valid for slow quenches and short times 
after the onset of spinodal instability.  In this linear model, the size of 
correlated domains scales with the quench rate $\tau$ to the same 
power as observed in the underdamped numerical simulations and predicted by the ``freeze-out'' proposal.  This provides 
analytical evidence for the ``freeze-out'' hypothesis.  A similar analytic model was introduced in the
context of a 1+1 dimensional classical condensed matter system in
\cite{Jacek}          

In contrast to previous approaches to the problem of defect formation in 
quantum field theory, we have allowed for the process of back reaction, 
which permits the quantum field to exit the region of spinodal instability.  
The power law exponent is the same whether it is measured at very early times with the analytic linear model (Fig. \ref{fig-apowerlaw})
or, in the numerical simulations, at early times in the middle of the spinodal region at the minimum value of $M_{eff}^2$ 
(Fig. \ref{fig-eslowpowerlaw}) or, at 
late times in the stable region at the maximum value of $M_{eff}^2$ (Fig. \ref{fig-lslowpowerlaw}).  
This is consistent with the idea that the relevant processes for the growth 
of domains occur near the critical point.  However, back reaction is 
{\it essential} to the freezing in of the value of the power law exponent and
to the ``freeze-out'' hypothesis. Although the power law is 
accurately recorded in the early time analytical 
model, if back reaction is ignored and the linear model is (incorrectly)
extrapolated to late times, the predicted scaling exponent is different

\begin {equation}
k_{max} \sim \tau^{-\frac {1} {2}}.
\end {equation}

The results derived from the numerical simulation of the mode function equation with back reaction 
and the analytical solution around the critical point are evidence supporting 
the first verification of the ``freeze-out'' scenario in 3 spatial dimensions 
and in a realistic system relevant to the early Universe. It is at first 
surprising that arguments based on {\it classical equilibrium} critical 
scaling apply in the context of {\it dynamical} quantum field theory. 
A possible explanation rests with the high temperature initial conditions 
and the properties of classical $\lambda\Phi^4$ theory in FRW spacetime. 
Since the initial temperature is much higher than the initial effective mass

\begin {equation}
T_0 >> m_{eff}(0) 
\end {equation} 

\noindent the field is approximately conformally invariant.  
In an FRW Universe, a conformally invariant field initially in thermal
equilibrium will remain in equilibrium at a redshifted temperature

\begin {equation}
\label {eq-temp}
T(t) = \frac {T_0} {a(t)}.
\end {equation}

\noindent A well-known example is the redshifting of the blackbody
spectrum of the cosmic microwave background radiation as the Universe expands.
To illustrate the approximate conformal invariance in our model we 
compare the dynamical effective mass from the simulations 
with the effective mass of a theory with the equilibrium redshifted 
temperature of Eq. (\ref{eq-temp})  

\begin{equation}
\label{eq-equilmass}
M^2_{eq}=-m^2+\frac {\lambda} {24} \frac { T^2_0} {a^2(t)}.
\end{equation}

\noindent The results are shown in Fig. \ref{fig-equilibrium}. Only after the onset of the
spinodal instability do the equilibrium and dynamical effective 
mass differ significantly.  The approximate conformal invariance of 
the theory means that the finite-temperature effective potential and the 
critical behavior derived from it are approximately valid until 
times near the onset of the spinodal instability.  The high temperature 
initial conditions also offer a possible explanation for the observed 
classical behavior of the system.  At such high temperature, thermal fluctuations 
dominate over quantum vacuum fluctuations. 

The agreement in the value of the power law exponent between the 
microscopic evolution equations of the quantum field theory and
phenomenological critical scaling supports the contention that 
quantum critical systems in the early Universe share, in certain circumstances, many of the 
properties of their classical counterparts.  It would be very interesting to further 
explore these connections.  To do so, however, it is necessary to go beyond the strong 
coupling and high temperature conditions used in this research.

The results derived in this paper represent a preliminary step in a 
first-principles approach to the calculation of the 
topological defect density immediately following the completion of a 
second-order phase transition in the early Universe.
Our work uses a microscopic quantum field theory and incorporates 
realistic initial conditions and a physical quench mechanism.  However, the 
domain wall defects formed in this model are inconsistent with 
cosmological observations. It is not difficult in principle to apply 
the methods used in this paper to more cosmologically realistic theories
such as cosmic string models.  The analytical and numerical evidence for the 
power law scaling of the domains is expected to hold in a more realistic model.

The formalism of the 2PI-CTP effective action and related techniques can be used 
to probe more general questions related to the physical aspects of quantum 
critical dynamics \cite{Winn}. To observe and isolate quantum processes it is
necessary to relax the assumption of high temperature initial conditions, 
allowing quantum vacuum fluctuations to dominate over thermal fluctuations.  Also important are
the detailed properties of the system-bath interaction.  A system-bath interaction such as proposed in \cite{Cal} 
could address such issues as critical slowing down and the role of dissipation and noise.  Such work is in progress.

\section*{Acknowledgments}
This work is supported in part by the National Science Foundation under grant PHYS-9800967.  
GJS would like to thank the National Scalable Computer Project at the University of Maryland for access to computing
resources.

\newpage
\begin{figure}[htb]
\begin{center}
\epsfig{file=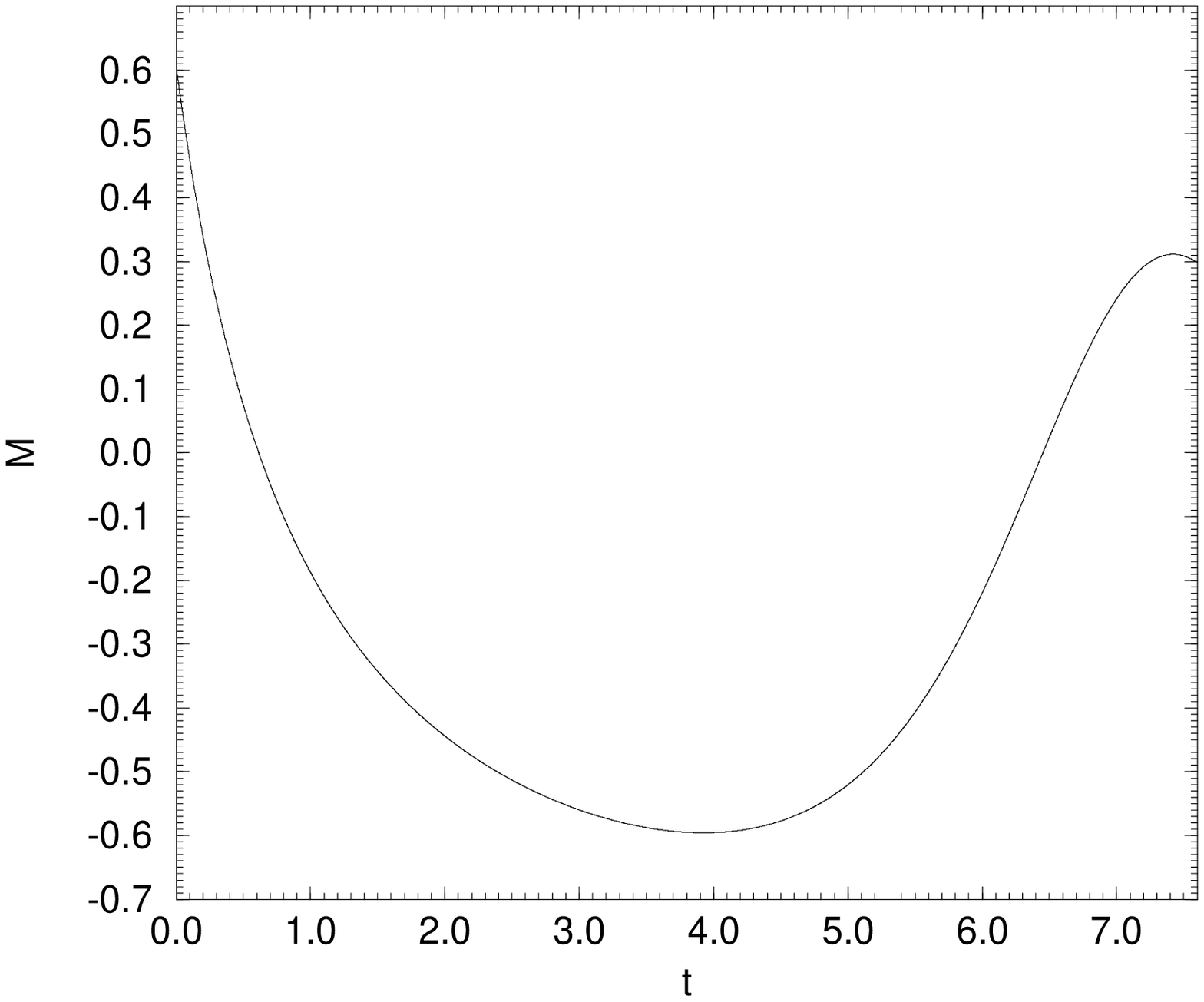, width=4.0 in}
\end{center}
\caption{Plot of the square of the (renormalized) effective mass 
$M=m^2_{eff}$ vs. cosmological time $t$ for quench parameter 
$\tau=1.0$.} 
\label{fig-meff}
\end{figure}

\begin{figure} [htb]
\begin{center}
\epsfig{file=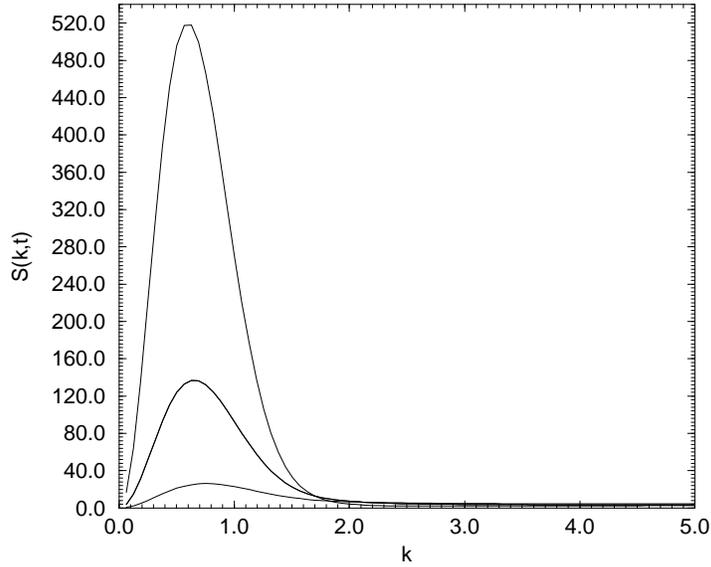, width=4.0 in}
\end{center}
\caption{Plot of the Fourier-space structure factor, $S(k,t)=k^2G(k,t)$, vs. 
comoving momentum $k$ at various times during the evolution for quench parameter $\tau=1.0$.  The bottom 
curve is a snapshot of $S(k,t)$
at time $t=3.9$. The middle curve is a snapshot at $t=5.4$.  The top curve is a snapshot at $t=6.9$.}
\label{fig-structure}
\end{figure}

\begin{figure} [htb]
\begin{center}
\epsfig{file=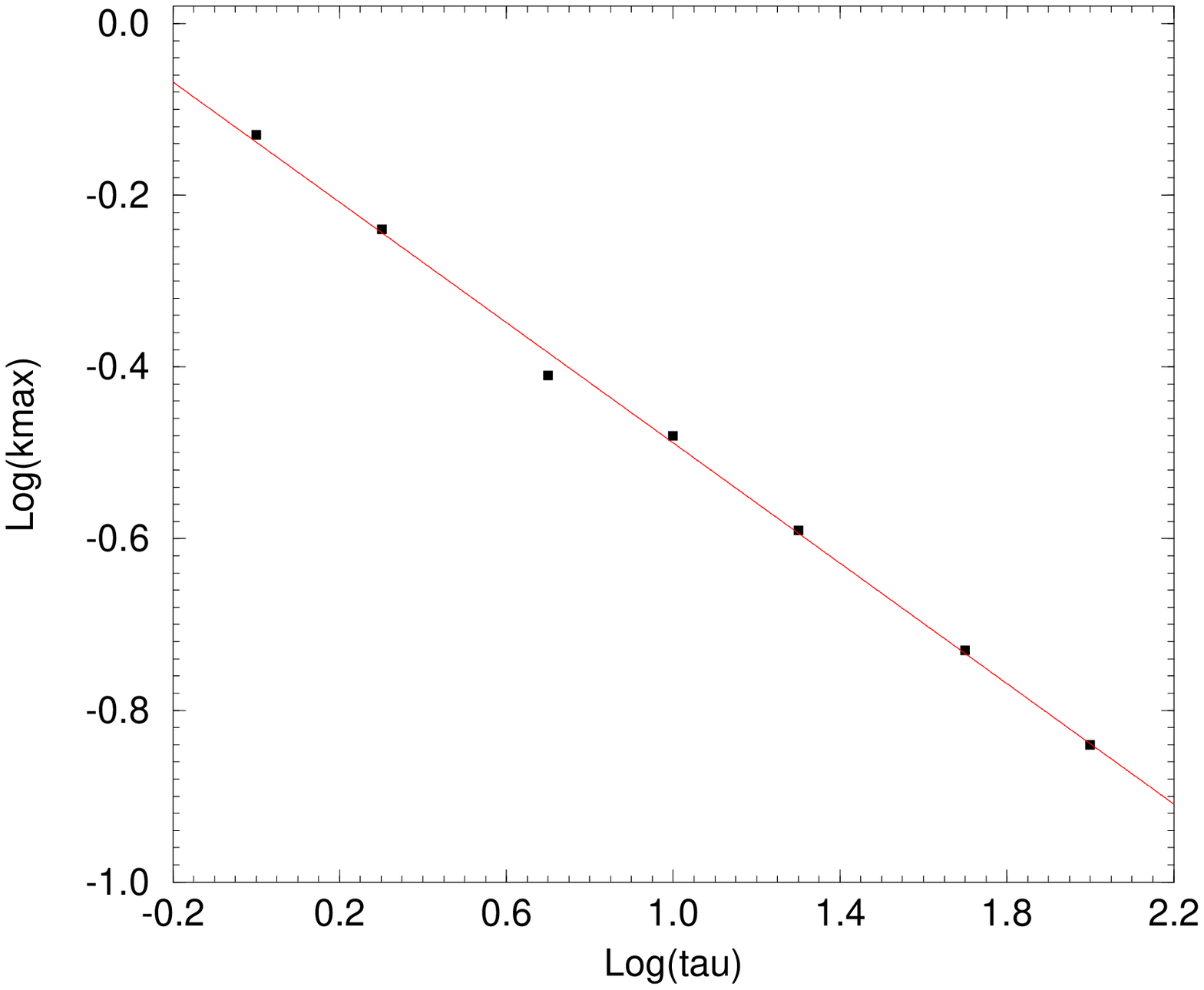, width=4.0 in}
\end{center}
\caption{Plot of $\log_{10}(k_{max})$ vs. $\log_{10}(\tau)$ in the case of
slow, underdamped quenches ($\tau \geq 1.0$) for domains formed early in the phase transition, at the time when the square
of the effective mass reaches a minimum value. 
Filled squares are measurements from the numerical simulations. The solid
line is a plot of the best-fit linear function to the data, 
$\log_{10}(k_{max})=-0.35\log_{10}(\tau)-0.14$.}
\label{fig-eslowpowerlaw}
\end{figure}

\begin{figure} [htb]
\begin{center}
\epsfig{file=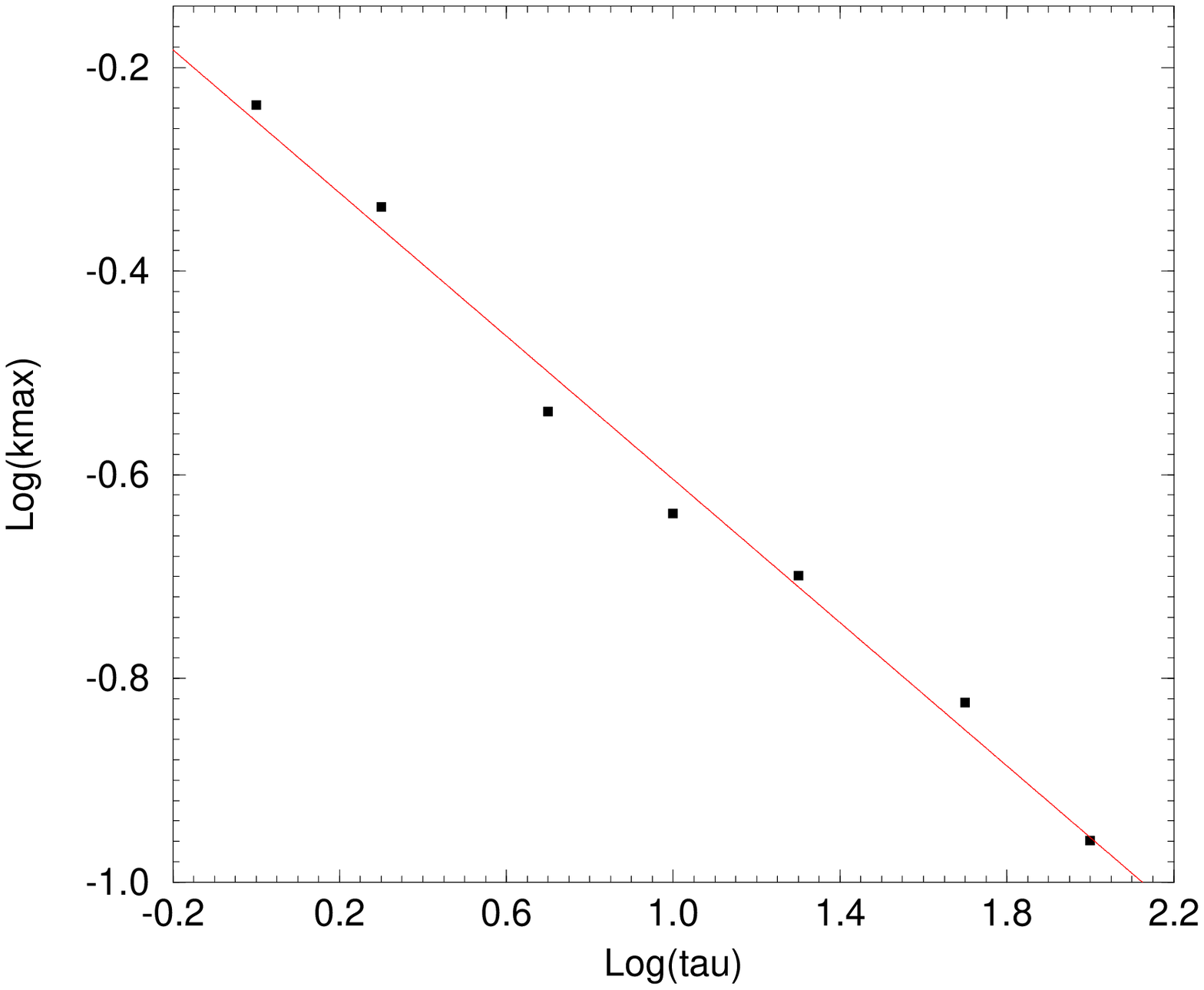, width=4.0 in}
\end{center}
\caption{Plot of $\log_{10}(k_{max})$ vs. $\log_{10}(\tau)$ in the case of slow,
underdamped quenches ($\tau \geq 1.0$) for
domains formed late in the phase transition, at the time when the square
of the effective mass reaches a local maximum value.  Filled squares are
measurements from the numerical simulations. The solid
line is a plot of the best-fit linear function to the data, 
$\log_{10}(k_{max})=-0.35\log_{10}(\tau)-0.26$.}
\label{fig-lslowpowerlaw}
\end{figure}

\begin{figure} [htb]
\begin{center}
\epsfig{file=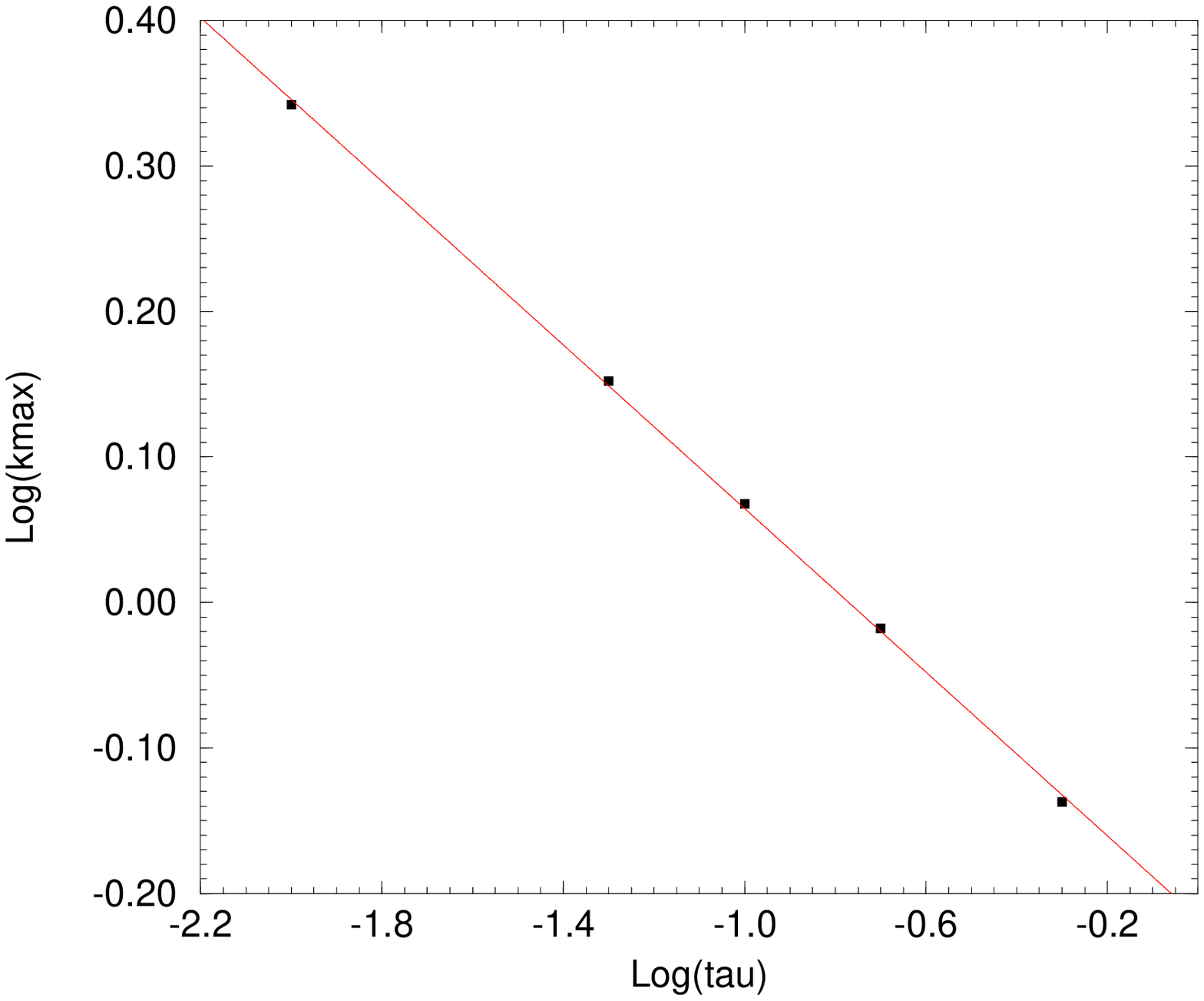, width=4.0 in}
\end{center}
\caption{Plot of $\log_{10}(k_{max})$ vs. $\log_{10}(\tau)$ in the case of fast,
overdamped quenches ($\tau <1.0$) for
domains formed late in the phase transition, at the time when the square
of the effective mass reaches a local maximum value.  Filled squares are
measurements from the numerical simulations. The solid
line is a plot of the best-fit linear function to the data, 
$\log_{10}(k_{max})=-0.28\log_{10}(\tau)-0.22$.}
\label{fig-lfastpowerlaw}
\end{figure}

\begin{figure} [htb]
\begin{center}
\epsfig{file=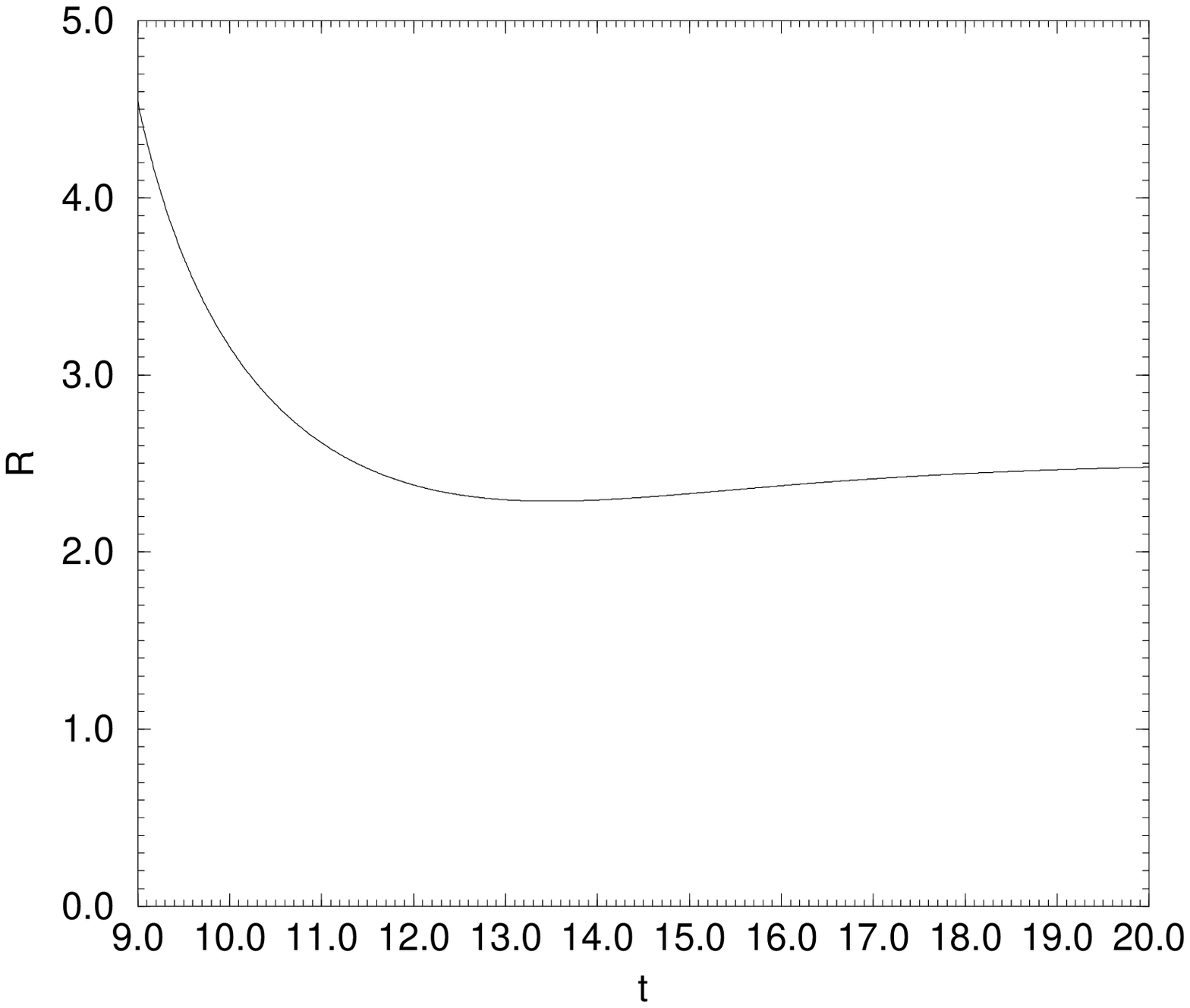, width=4.0 in}
\end{center}
\caption{Plot of the ratio, $R(t)=\frac {k_{max}(t)} {\rho(t)}$, 
vs. cosmological time $t$ for quench parameter
$\tau=10.0$.  The defect density $\rho(t)$ was calculated using the analytic modes, Eq.(\ref{eq-amode}), and the
Halperin-Mazenko-Liu formula, Eq. (\ref{eq-density}), with a cutoff at the
maximum momentum of the unstable band. 
The maximum $k_{max}(t)$ in the structure function, $S(k,t)=k^2G(k,t)$, 
was also determined using the analytic modes, Eq.(\ref{eq-amode}). At late times
this ratio approaches a constant that is independent of the quench 
parameter.}
\label{fig-ratio}
\end{figure}

\begin{figure} [htb]
\begin{center}
\epsfig{file=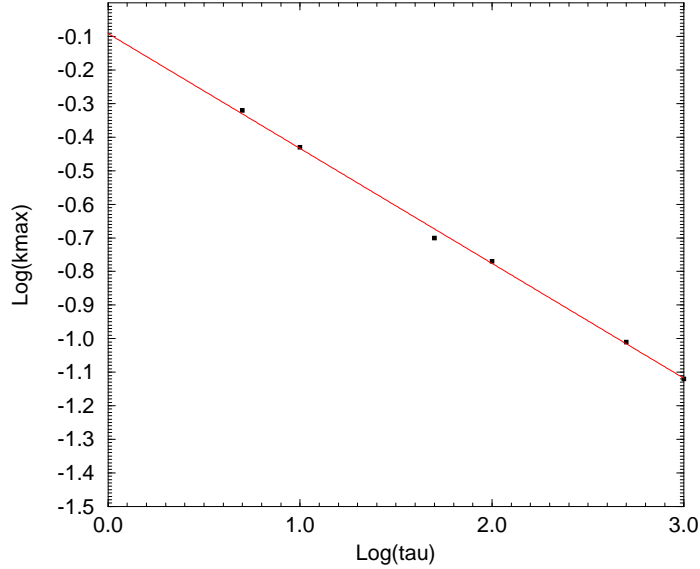, width=4.0 in}
\end{center}
\caption{Plot of $\log_{10}(k_{max})$ vs. $\log_{10}(\tau)$ in the analytic model, Eq. (\ref{eq-amode}).  The maximum,
$k_{max}$, was determined as soon as a peak was evident in the structure function, $S(k,t)$. Filled squares are 
measurements from plots of the analytic modes.  The solid line is a plot of the best fit linear function to the 
data, $\log_{10}(k_{max})=
-0.34\log_{10}(\tau)-0.09$.}
\label{fig-apowerlaw}
\end{figure}

\begin{figure} [htb]
\begin{center}
\epsfig{file=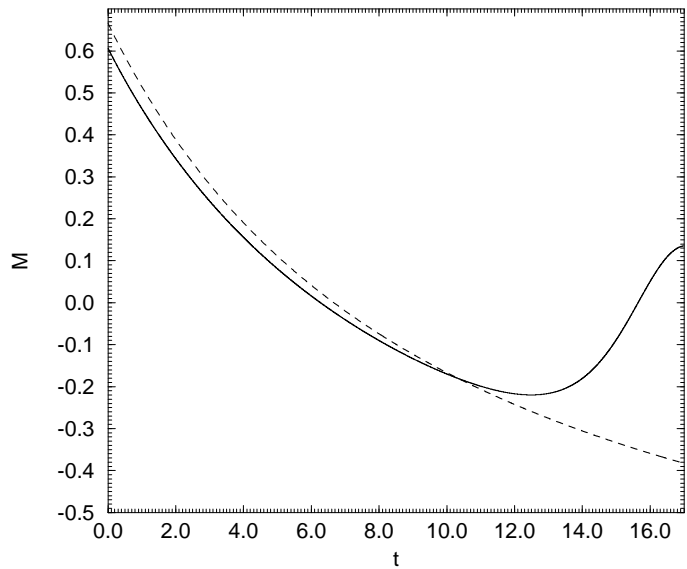, width=4.0 in}
\end{center}
\caption{Plot of the square of the effective mass $M=m^2_{eff}$ 
vs. cosmological time $t$ in the equilibrium 
case (dashed line) where
$m^2_{eff}=-m^2+\frac {\lambda} {24} \frac { T^2_0} {a^2(t)}$ and the in 
nonequilibrium case (solid line) where $m^2_{eff}$ is determined by
the full numerical simulations.  The slight difference in the equilibrium and nonequilibrium curves at 
the beginning of the evolution is due to the difference between the general Hartree-Fock effective mass given by Eq. (\ref{eq-minit}) and 
the high temperature, small $\lambda$ limit given by Eq. (\ref{eq-highTm}). }
\label{fig-equilibrium}
\end{figure}

\end{document}